# Resistivity of Non-Fermi Liquid $U_2Pt_2In$ Under Pressure

P. Estrela[a,*], T. Naka[a,b], A. de Visser[a], F.R. de Boer[a], L.C.J. Pereira[c], M. Almeida[c]

[a] *Van der Waals-Zeeman Institute, University of Amsterdam, Valckenierstraat 65, 1018 XE Amsterdam, The Netherlands*
[b] *National Research Institute for Metals, 1-2-1 Sengen, Tsukuba, Ibaraki 305-0047, Japan*
[c] *Dep. Química, Instituto Tecnológico e Nuclear, Apartado 21, 2686-953 Sacavém, Portugal*

**Abstract**

Non-Fermi liquid behaviour in single-crystalline $U_2Pt_2In$ has been studied by means of resistivity experiments ($\mathbf{I}\|\mathbf{c}$) under hydrostatic pressure ($P<1.5$ GPa). At ambient pressure the resistivity $\rho(T)$ follows a power law $\rho \sim T^\alpha$ with $\alpha \sim 0.5$. Upon applying pressure $\alpha$ increases. For $P>1$ GPa a minimum develops in $\rho(T)$. A study of the field dependence of the minimum confirms its magnetic origin. The ratio $c/a$ is proposed as the effective control parameter, rather than the unit cell volume.

The intermetallic compound $U_2Pt_2In$ (tetragonal $Zr_3Al_2$-type structure) is one of the few stoichiometric compounds known to exhibit non-Fermi liquid (NFL) properties [1] at ambient pressure, as evidenced by a pronounced $c/T \sim -\ln T$ contribution to the specific heat below $\sim 6$ K [2] and a low temperature electrical resistivity described by a power law $\rho \sim T^\alpha$ with $\alpha \sim 1$ for $\mathbf{I}\|\mathbf{a}$ and $\alpha \sim 0.5$ for $\mathbf{I}\|\mathbf{c}$ [3]. A broad maximum at $T_{max}=8$ K in the magnetization measured for $\mathbf{B}\|\mathbf{c}$ [3] and a rapid increase below $T_{max}$ of the line width of a dynamic component in the zero-field μSR spectra [2] indicate the presence of short-range magnetic correlations. μSR experiments confirm the absence of (weak) magnetic order. Although the residual resistivity of $U_2Pt_2In$ is quite high ($\rho_0 \sim 100$ μΩcm), X-ray [3] and neutron diffraction [4] experiments carried out on single-crystalline samples rule out the possibility of significant crystallographic disorder (namely intersite exchange). The location of $U_2Pt_2In$ at the border line of magnetic and non-magnetic compounds in a Doniach-type of diagram for the $U_2T_2X$ family [5] suggests the proximity to a quantum critical point as the origin of the non-Fermi liquid behaviour in this compound.

Recently, polycrystalline samples of $(U_{1-x}Th_x)_2Pt_2In$ were prepared for $x \leq 0.1$ [6] and studied by magnetization and μSR, which showed that Th doping stabilizes the magnetic state in $U_2Pt_2In$ [7]. Assuming that the main role of Th doping is an expansion of the unit cell, applying hydrostatic pressure gives the opportunity to study the effects of a unit cell volume reduction and an eventual recovery of the Fermi liquid state.

---
[*] Corresponding author. Fax: +31-20-5255788; e-mail: estrela@wins.uva.nl

Resistivity measurements in the temperature range 0.3-300 K, under pressures up to 1.5 GPa and magnetic fields up to 8 T, were carried out in bar-shaped single-crystals of $U_2Pt_2In$ using a standard four-probe method. A CuBe piston cylinder-type clamp cell was used with Fluorinert acting as pressure transmitting medium. The pressure values were corrected for an empirical efficiency of 80%.

The zero-pressure resistivity curve for **I**∥**c** follows a power law of the type $\rho \sim T^\alpha$ with $\alpha \sim 0.5$ in the temperature range $0.3 \leq T \leq 2.3$ K [3]. Upon increasing the pressure, $\alpha$ gradually increases for $P<1$ GPa (fig.1). However, as the pressure is increased further, a minimum in the resistivity develops, which shifts to higher temperatures and becomes more pronounced at higher pressure values: $T_{min} \sim 1.2$ K at $P=1.2$ GPa and $T_{min} \sim 2.1$ K at 1.45 GPa.

In order to investigate whether the observed minimum has a magnetic origin, measurements of $\rho(T)$ under magnetic field (**B**∥**I**) were performed at $P=1.45$ GPa (fig.2). $T_{min}$ decreases smoothly with the strength of the applied magnetic field: $T_{min} \sim 1.7$ K for $B=4$ T and $T_{min} \sim 1.2$ K for 8 T. Also the minimum becomes less pronounced with field, and is almost suppressed at 8 T. Due to the limited range of temperatures and fields available and the relative uncertainty in determing $T_{min}$ (especially for the higher fields where the minimum is very shallow), it is not possible at the moment to clearly trace the evolution of $T_{min}$ with $B$. The inset on fig.2 shows that the data can be equally fitted with a linear dependence ($B = B_0(1 - T/T_0)$ with $B_0 = 17.0 \pm 1.1$ T and $T_0 = 2.20 \pm 0.05$ K) or a typical antiferromagnetic-like field dependence of the ordering temperature: $B = B_0[1 - (T/T_0)^2]^\beta$ with $B_0 = 9.8 \pm 1.1$ T, $T_0 = 2.14 \pm 0.03$ K and $\beta = 0.82 \pm 0.17$. The field effect on the resistivity curves strongly suggests that $T_{min}$ has a magnetic origin. Whether $T_{min}$ is associated with magnetic ordering remains undecided at the moment.

The structure in $\rho(T)$ at $P=1.45$ GPa resembles in part the one of an antiferromagnetic phase transition of the spin-density wave type, as observed e.g. in the heavy-fermion antiferromagnets $URu_2Si_2$ ($T_N=14$ K) [8] and $U(Pt_{0.95}Pd_{0.05})_3$ ($T_N=5.8$ K) [9]. In these compounds $\rho(T)$ rises below $T_N$ because of the opening of the energy gap. At about $0.9T_N$ $\rho(T)$ develops a local maximum, where below the resistivity drops because of the ordered structure. The data in Fig.1 show that this local maximum is not observed at temperatures $>0.15T_N$, which might be related to the large residual resistance values of our samples.

The emergence of a magnetic component to $\rho(T)$ upon applying pressure is quite surprising. Applying pressure on a compound at the magnetic instability normally leads to an increase of the control parameter $J$ and brings the compound in the non-ordering Fermi-liquid regime. A possible explanation for this unusual behaviour might be offered by assuming that the control parameter $J$ is not governed by the volume, but by the $c/a$ ratio of the tetragonal unit cell. This is supported by the comparison of the unit cell volumes and $c/a$ ratios of $U_2Pt_2In$ and the antiferromagnets $U_2Pd_2In$ ($T_N=37$ K) and $U_2Pt_2Sn$ ($T_N=15$ K) [10]. Whereas the unit-cell volume of $U_2Pt_2In$ is smaller than the one of $U_2Pd_2In$ and larger than the one of $U_2Pt_2Sn$, the $c/a$ ratio is always smaller (for simplicity we have not taken into account the doubling of the crystal structure along the c-axis in single-crystalline $U_2Pt_2In$) [11]. Thus the appearance of magnetic ordering under pressure might be the result of a $c/a$ ratio increase. This in turn requires the compressibility to be anisotropic. X-ray diffraction experiments to determine the lattice parameters of $U_2Pt_2In$ under pressure and of $(U,Th)_2Pt_2In$ are underway to investigate this hypothesis.

In summary, we have measured the resistivity of single-crystalline $U_2Pt_2In$ under moderate pressures and conclude that $U_2Pt_2In$ is easily driven away from its non-Fermi-liquid state, as expected for compounds at a quantum critical point.

**Acknowledgements**:

P.E. acknowledges the European Commission for a Marie Curie Fellowship within the TMR programme.

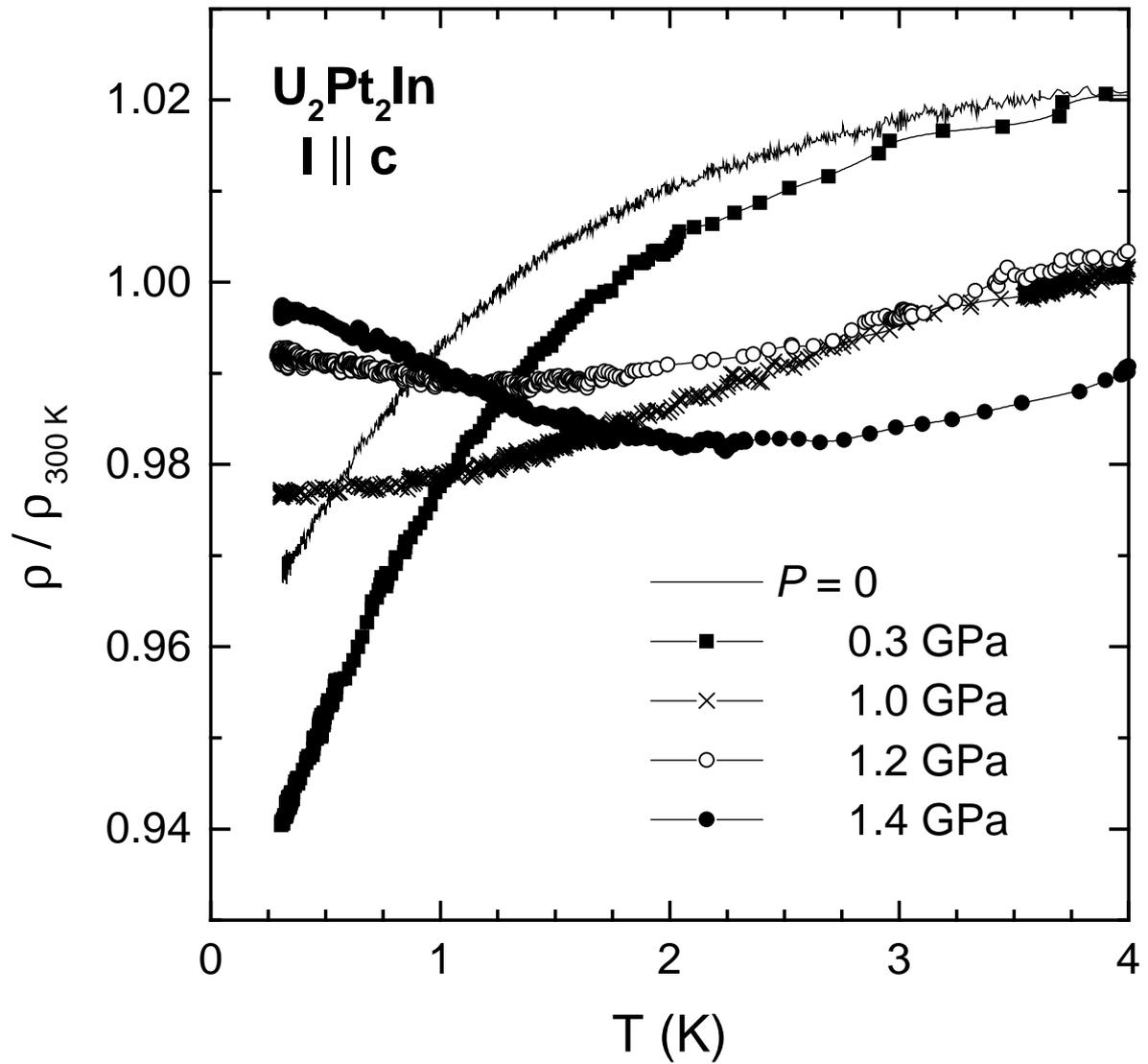

Fig. 1 - Temperature dependence of the electrical resistivity (normalized to the respective room temperature values) at several pressures for **I**||**c** in single-crystalline $U_2Pt_2In$.

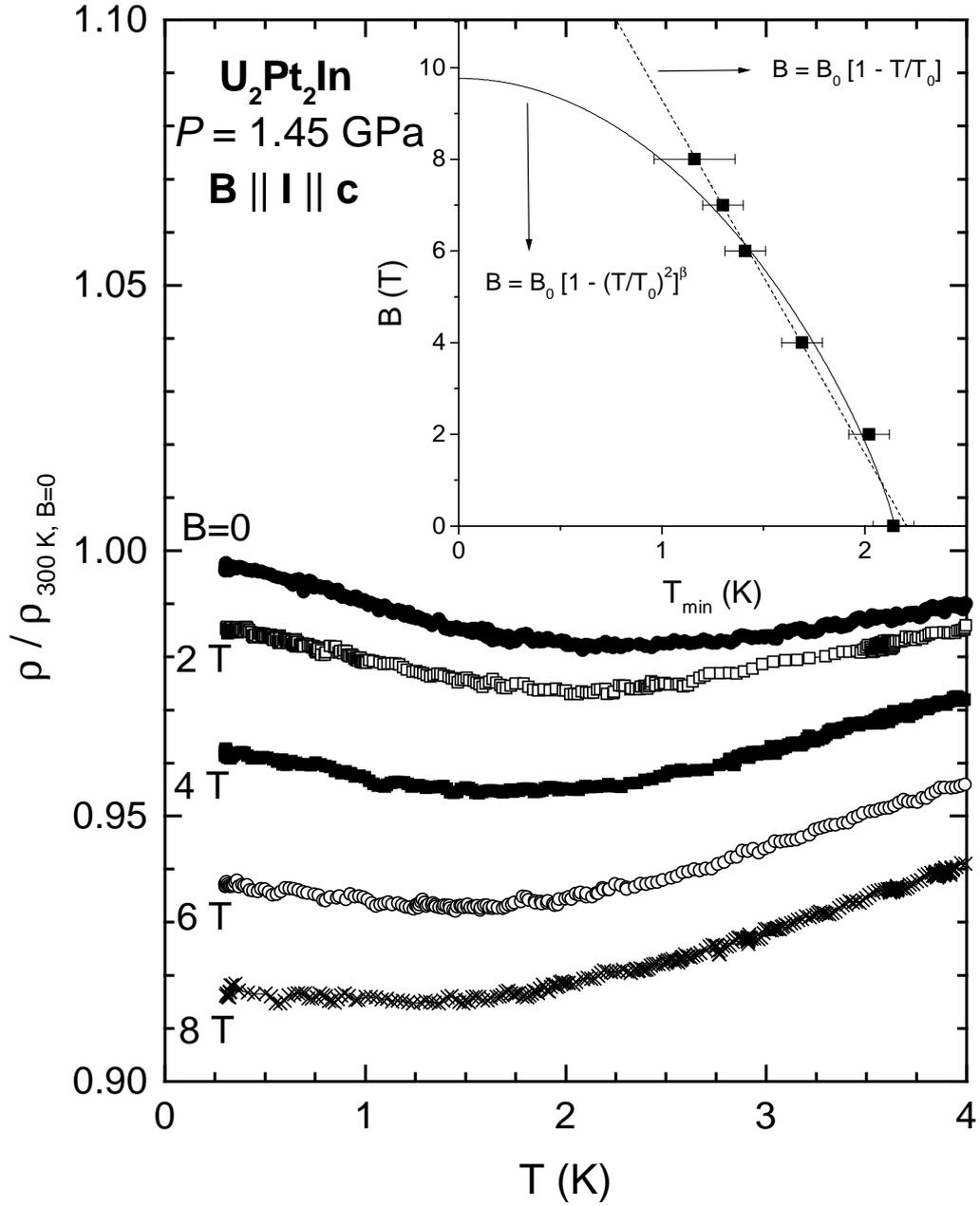

Fig. 2 - Normalized ρ(*T*) at *P*=1.45 GPa for several magnetic fields applied parallel to the current direction (**B**||**I**||**c**). Inset: field dependence of $T_{\min}$ (the lines are fits to the data).